\documentclass[conference]{IEEEtran}
\IEEEoverridecommandlockouts
\usepackage{cite}
\usepackage{amsmath,amssymb,amsfonts}
\usepackage{algorithmic}
\usepackage{graphicx}
\usepackage{textcomp}
\usepackage{xcolor}
\usepackage[hidelinks]{hyperref}
\usepackage{cleveref}
\usepackage{algorithm}
\usepackage{gensymb}
\usepackage{booktabs}
\usepackage{float}
\usepackage{graphicx}
\usepackage{subcaption}

\def\BibTeX{{\rm B\kern-.05em{\sc i\kern-.025em b}\kern-.08em
    T\kern-.1667em\lower.7ex\hbox{E}\kern-.125emX}}

\def \diag{\text{diag}}
    
\begin{document}

\title{Impacts of Dynamic Line Ratings on the \\
ERCOT Transmission System}

\author{\IEEEauthorblockN{Thomas Lee}
\IEEEauthorblockA{\textit{Institute for Data, Systems, \& Society} \\
\textit{Massachusetts Institute of Technology}\\
Cambridge, MA, USA \\
t\_lee@mit.edu}
\and
\IEEEauthorblockN{Vineet Jagadeesan Nair}
\IEEEauthorblockA{\textit{Dept. of Mechanical Engineering} \\
\textit{Massachusetts Institute of Technology}\\
Cambridge, MA, USA \\
jvineet9@mit.edu}
\and
\IEEEauthorblockN{Andy Sun}
\IEEEauthorblockA{\textit{Sloan School of Management} \\
\textit{Massachusetts Institute of Technology}\\
Cambridge, MA, USA \\
sunx@mit.edu}
}

\maketitle

\begin{abstract}
Grid regulators and participants are paying increasing attention to Dynamic Line Ratings (DLR) as a new approach to address transmission system bottlenecks. In this paper, a thorough comparison of DLR, Ambient Adjusted Ratings (AAR), and the traditional Static Line Ratings (SLR) are conducted on a synthetic ERCOT grid. Estimates of DLR and AAR are calculated using an equation based on heat balance physics, along with high-resolution weather data of temperature and wind velocities. A constraint generation method for contingency screening is developed for solving security-constrained optimal power flow. Numerical results suggest that employing DLR could double the benefits compared to those of AAR relative to SLR, in terms of system costs, renewable curtailment, and emissions.

\end{abstract}

\begin{IEEEkeywords}
Dynamic line ratings, transmission capacity, security-constrained optimal power flow, reliability.
\end{IEEEkeywords}

\section{Introduction and Motivation} 
Power grids in the US face serious challenges with their transmission networks. Transmission projects are capital intensive and can take up to 5-10 years to site, permit, and build. Furthermore, regulatory barriers prevent many new projects from ever seeing the light of day. Such a shortage of transmission capacity causes grid congestion and greater curtailment of renewables such as wind and solar. This problem is only likely to get worse in the future, as the interconnection capacity queue has already surpassed 1,300 GW in 2021 and continues to grow \cite{rand2022queued}. As a way to make the best use of existing transmission capacities, FERC ruled in 2021 that all transmission providers should use Ambient Adjusted Ratings (AAR). In 2022, FERC is further investigating the usage of Dynamic Line Ratings (DLR). This project explores the potential of DLR to provide flexible transmission capacity and optimize the use of existing infrastructure, in order to reduce the need for transmission expansion.

\section{Background and Contributions} 

To motivate the need for DLR, we provide a brief overview on transmission line ratings. The main safety and reliability reasons for imposing line flow limits in terms of maximum current or apparent power flow are summarized below \cite{DepartmentofEnergy2019Dynamic2019}:
\begin{enumerate}
    \item Prevent thermal annealing of the conductor metal, which can damage strength and electrical properties.
    \item Maintain vertical clearance of transmission lines to the ground and vegetation, by preventing line sagging.
    \item Preserve stability limits, which can be translated to power flow limits for dispatch processes that do not handle transient timescales.
\end{enumerate}

\subsection{Types of line ratings}
\begin{enumerate}
    \item \textbf{Static Line Ratings (SLR)} are conservative line rating estimates based on average or worst-case values for assumed weather conditions, commonly used by most grid operators for safety and reliability reasons.
    
    \item \textbf{Ambient Adjusted Ratings (AAR)} are adjusted from SLR using only ambient air temperature. PJM and ERCOT already implemented AAR based on temperature lookup tables, which are coarse and crude approximations \cite{Kolkmann2019ManagingRatings}. In 2021, FERC issued Order No. 881, requiring all transmission providers to use AAR \cite{FERNOI:online}.

    \item \textbf{Dynamic Line Ratings (DLR)} are real-time line ratings that take into account prevailing environmental conditions such as ambient air temperature and wind as well as conductor status measurements (from low-cost installed sensors) such as circuit current, line tension and sag, with a fine spatial-temporal resolution. As a result, DLR is less conservative than SLR and AAR and allows grid operators to utilize unused line rating margins to increase power transfer and reduce grid congestion \cite{Murphy2021DynamicOverview, FernandezReviewIntegration}. In 2022, FERC issued a notice of inquiry regarding the usage of DLRs \cite{FERNOI:online}.
\end{enumerate}

\subsection{Prior work on line ratings}
Comprehensive reviews of DLR and overhead line monitoring technologies, including real-world pilot DLR implementations, are given in  \cite{FernandezReviewIntegration, erdincc2020comprehensive}. Since high wind speeds simultaneously influence both wind energy generation and the thermal convection affecting DLR, the application of DLR for wind power integration is specifically examined in \cite{FernandezReviewIntegration}.

Compared with static ratings, DLR introduces new risks and uncertainties. Recent studies use machine learning to develop operational forecasts of DLR values \cite{sobhy2021overhead},\cite{aznarte2016dynamic}. Recent optimization approaches seek to incorporate DLR uncertainties while balancing risk and efficiency; for example, \cite{wang2018risk} develops a distributionally robust optimization model with an uncertainty set capturing the correlation between wind generation and line ratings, \cite{Morrow2014ExperimentallyRating} uses partial least squares regressions, and \cite{dupin2019optimal} uses bi-level stochastic optimization to set DLR values while explicitly penalizing risk-based probabilistic DLR forecasts and the grid operator's risk aversion. Recent works like \cite{Racz2022PerformanceExperiences} have also analyzed the performance, accuracy, and reliability of real-world small-scale DLR demonstration projects.

\subsection{Contributions}
While previous studies only focus on DLR relative to SLR, this paper adapts a multiplicative DLR formula in order to explicitly quantify the comparative impacts between AAR and DLR. This is a very pertinent policy question for FERC and regions such as ERCOT which already have AAR but may consider future DLR. In contrast to some studies which focus on individual lines or local regions, this paper implements a method to efficiently simulate realistic system operations when adopting widespread DLR across an entire network.

\section{Methodology}
\subsection{DLR estimation model} 
In order to estimate AAR and DLR values, an existing model proposed in literature \cite{wallnerstrom2014impact} is modified to calculate the multiplicative factor for line ampacity using wind velocity and ambient air temperature. In particular, we have:
\begin{subequations}
\begin{align}
&\eta(v, T) =\frac{I_{\mathrm{max}}^{D L R}}{I_{\max }^{\mathrm{SLR}}} \approx \eta_{v}  \eta_{T}, \\
&\eta_{v} =\sqrt{\frac{K_{angle}}{K_{angle}^{SLR}}} \cdot \frac{v^{0.26}}{v_{\mathrm{SLR}}^{0.26}}\max\{1,0.566\cdot(\frac{\rho_{f}}{\mu_{f}}Dv)^{0.04}\}, 
\end{align}\label{dlr}
\end{subequations}
where $K_{angle}^{SLR} = 0.388$ when $\phi^{SLR} = 0$ in the worst case scenario ($\phi = 90 \degree$ is the ideal condition since the convective heat transfer loss is maximized when the wind is perpendicular to the power lines). $\eta$ is the overall DLR capacity increase factor while $\eta_v$ and $\eta_T$ correspond to the DLR factors based only on either wind velocity or temperature (AAR).

One key distinction from \cite{wallnerstrom2014impact} is that we incorporate the actual wind direction rather than assuming it to always be perpendicular to the line. This is accounted through the angle factor $K_{angle}$ computed using the IEEE standard 738-2012 \cite{ieee2012ieee} as shown below, where latitudes and longitudes were converted to $(x,y)$ coordinates using the Universal Transverse Mercator (UTM) projection.
\begin{equation*}
\begin{aligned} 
    & \text{Wind speed} = v = \sqrt{v_x^2 + v_y^2}, \; \phi = \theta_{wind} - \theta_{cond}\\
    & \text{Wind angle} = \theta_{wind} = \tan^{-1}\left(\frac{v_y}{v_x}\right), \\
    & \text{Conductor angle} = \theta_{cond} = \tan^{-1}\left(\frac{y_{to} - y_{from}}{x_{to} - x_{from}}\right), \\
    & K_{\text {angle }} = 1.194- \cos (\phi) +0.194 \cdot \cos (2 \phi) +0.368 \cdot \sin (2 \phi)
\end{aligned}
\end{equation*}
where $v_x, v_y$ are the horizontal and vertical components of the wind velocity and $v$ is the wind speed. $(x_{from}, y_{from})$ and $( x_{to}, y_{to})$ are the coordinates of the from and to buses of each branch, respectively. We also extend \cite{wallnerstrom2014impact} by using higher resolution (hourly) data and applying DLR to a much larger test system and running security-constrained economic dispatch to account for contingencies.

In accordance with \cite{wallnerstrom2014impact}, we make some simplifying assumptions. We assume that the temperature and wind speed independently affect the line ampacity. Further, we assume that the thermal and electrical properties such as the dynamic viscosity ($\mu_f$) and density ($\rho_f$) of air and the resistance of the conductor do not vary with temperature. The fixed ambient temperature ($T^{SLR}_A$), wind speed ($v^{SLR}$) assumed for the SLR calculation are set based on average values used by ERCOT. In the absence of line diameter data, we convert the available MVA thermal ratings to ampacity, and then leverage the strong empirical linear relationship between ampacity and diameters.

In the absence of real-time data on conductor currents or temperatures ($T_C$), we assume a high fixed value for the maximum allowable conductor temperature ($T_C = 100 \degree C$), which is equal to or higher than the normal conductor temperatures assumed by most transmission and distribution service providers (TDSP) in Texas\footnote{We obtained this data from the ERCOT Steady State Working Group Manual \url{https://www.ercot.com/committees/ros/sswg}}. This is a conservative and valid approach, since the  temperature factor $\eta_T$ decreases with respect to $T_C$, as the first-order derivative
\begin{align*}
    & \frac{\partial}{\partial T_C}\left(\frac{\sqrt{T_C - T_A}}{\sqrt{T_C - T_A^{SLR}}}\right) = \frac{T_A - T_A^{SLR}}{2 \sqrt{T_C - T_A}(T_C - T_A^{SLR})^{\frac{3}{2}}} 
\end{align*}
is negative as long as $T_A < T_A^{SLR}$, which generally holds true since $T^{SLR}_A$ is usually set by operators to a high value so as to be conservative, i.e. $T_A < T_A^{SLR} < T_C$. Thus, assuming a fixed high $T_C$ would give us a safer underestimate of the DLR factor.

In summary, a key advantage of \Cref{dlr} is its multiplicative decomposition into temperature-based and wind-based factors.
\begin{enumerate}
    \item \textbf{AAR} multipliers are calculated to equal temperature-only DLR ratios $\eta_T$, without using a floor, as $AAR = SLR\cdot \eta_T$, which approximates the current ERCOT practice of AAR. 
    \item \textbf{DLR} are calculated as $DLR = AAR \cdot \max\{1, \eta_v\}$. We assume that introducing full DLR should not worsen the ratings below the temperature-only AAR, which are currently operationally acceptable.
\end{enumerate}


AAR/DLR multipliers are applied only to short and medium-length lines below 100 km, because at longer line lengths voltage and stability limits become more stringent than thermal limits \cite{DOE:online}. Transformers are similarly kept at their static ratings. Of the 3,206 branches in the synthetic ERCOT grid, 72\% are sub-100 km lines (eligible for AAR/DLR), 1\% are longer lines, and 27\% are transformers.


Contingency line ratings represent the allowable branch flows during a contingency event. These values are not directly provided by the synthetic grid data. Instead, we use the line rating tables published by PJM \cite{pjmratings:online} to calculate the ratios between contingency and normal line ratings for all lines. Then, they are averaged to estimate a typical ratio of 1.146. 

\subsection{SC-DCOPF using contingency screening} 
This paper studies a lossless DCOPF model with N$-$1 preventative security constraints on transmission line contingencies, which reflects the ISO market clearing process. For simplicity, time-coupling constraints including ramp rates and unit commitment are ignored in this paper. A constraint generation approach, inspired by and simplified from \cite{weinhold2020fast}, is implemented to efficiently screen for a subset of active contingencies out of the combinatorially many possible monitored-contingency tuples. The pseudo-code is given in Algorithm \ref{alg:constrgen}. Details are described below.
First, for each dispatch hour $t$, an initial DCOPF base case model given below is solved without contingencies, where the time-dependent branch flow limits $\bar{\textit{\textbf{f}}}(t)$ are calculated using the DLR methodology described in the previous section.
\begin{align*}
\min _{P_{G}(t)} \quad & \sum_{i \in G} c_{i}\left(P_{G i}(t)\right) \\
\text{s.t.} \quad & \sum_{i=1}^N(P_{G i}(t)-P_{D i}(t))=0,\\
& P_{G i}^{\min } \leq P_{G i}(t) \leq P_{G i}^{\max }, \;\;\forall i \in G, \\
& - \bar{\textit{\textbf{f}}}(t) \leq \textbf{PTDF}(P_G - P_D) \leq \bar{\textit{\textbf{f}}}(t),
\end{align*}
where $G$ is the set of generators, $P_{Gi}$ is the real power output of generator $i$, $P_{Di}$ is the real power consumption of load $i$, and \textbf{PTDF} represents the power transmission distribution factors matrix. These pre-contingency flows are multiplied by the line outage distribution factors (LODF), an $L\times L$ matrix, to obtain hypothetical post-contingency flows. Here it is assumed that the rows of the LODF matrix correspond to monitored branches and columns correspond to contingency line outages. The $L\times L$ matrix $F^{cont}$ adds base case flows $f^{base}$ to outage-redistributed flows, whose element $F^{cont}_{b,c}$ represents line $b$'s flow upon outage $c$, as shown in step 2 in \cref{alg:constrgen}.
The $F^{cont}$ flows can be checked against the contingency line ratings to find active contingency violations. For each identified line-contingency pair $(b,c)$, append its corresponding contingency-PTDF to the active set of PTDF rows. The original (non-contingency) PTDF matrix is combined with the active set of contingency-PTDF rows and then solved with an updated PTDF formulation of DCOPF. Here violations of the branch limits are allowed by introducing new slack variables and penalized at \$2,000/MWh, corresponding to ERCOT's administratively set transmission shadow price cap for non-resolvable constraints. Then the resulting flows are checked again for new violations, and the procedure is repeated until convergence. During the numerical study, almost all hourly cases converge within 1-3 iterations in the while loop.
\begin{algorithm}[htbp]
\caption{Algorithm for preventative (N-1) SC-DCOPF}
\begin{algorithmic}[1]
\renewcommand{\algorithmicrequire}{\textbf{Input:}}
\renewcommand{\algorithmicensure}{\textbf{Output:}}
\REQUIRE \textbf{PTDF}, \textbf{LODF}, $P_{D}, P_{max}, \bar{f}^A, \bar{f}^B$\\
\ENSURE  $P_G^*$ and constraint shadow prices \\
\STATE Solve base case DCOPF $\to P_G^{base}, f^{base}$
\STATE Compute $F^{cont} = f^{base}\cdot\textbf{1}_{(1\times L)} + \textbf{LODF}\cdot \diag(f^{base}))$
\STATE Find violated contingencies $V = \{(b,c) : |F^{cont}_{b,c}| > \bar{f}^B_b\}$
\STATE Initialize $P^0\leftarrow \textbf{PTDF}$, $r^0\leftarrow \bar{f}^A$, $i \leftarrow 0$
\WHILE {$|V|>0$}
\STATE $i \leftarrow i+1$
\FOR {$(b,c) \in V$}
\STATE $P^i \leftarrow \begin{pmatrix} P^{i-1} \\ \textbf{PTDF}_b + \textbf{PTDF}_c\textbf{LODF}_{b,c} \end{pmatrix}$, $r^i \leftarrow \begin{pmatrix} r^{i-1} \\  \bar{f}^A_b \end{pmatrix}$
\ENDFOR
\STATE Solve DCOPF($P^i,r^i$) $\to P_G^*$
\STATE Update $f^{base} \leftarrow \textbf{PTDF}(P_G^* - P_D)$
\STATE Update $F^{cont}$ as line 2 and $V$ as line 3
\ENDWHILE
\RETURN $P_G^*$ and solution
\end{algorithmic}\label{alg:constrgen}
\end{algorithm}

\subsection{Parallelization} 
Incorporating contingencies increases the computational burden, so hourly optimization calculations are parallelized, as the SC-DCOPF model does not involve inter-temporal constraints such as unit commitment and ramping. This parallelization can significantly speed up the computation as shown in the next section.

\section{Computational Experiments and Results}
\subsection{Weather data} 
\begin{figure}[htbp]
     \centering
    \includegraphics[width=0.49\columnwidth]{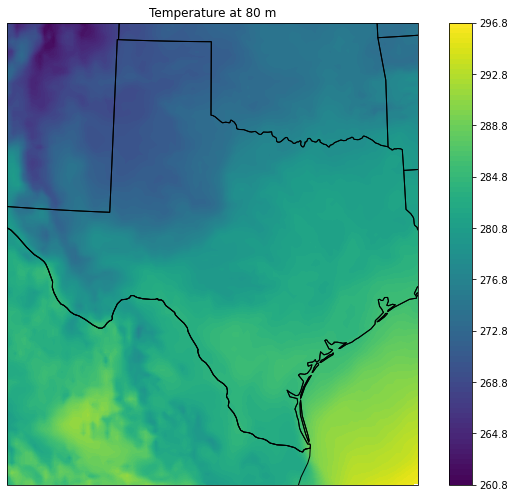}
    \includegraphics[width=0.49\columnwidth]{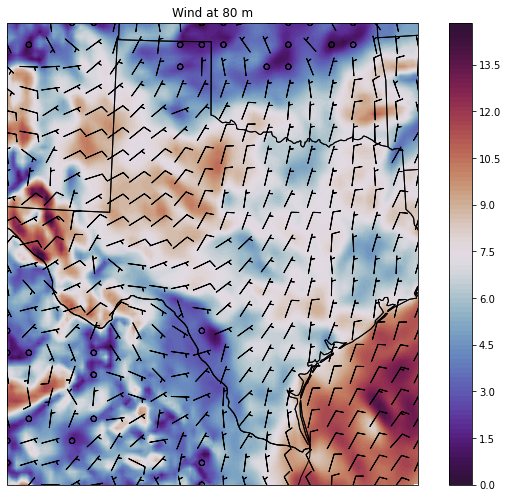}
        \caption{Weather data for 1AM on January 1, 2016. Temperature (in K). Wind speed (m/s) and direction.}
        \label{fig:weather}
\end{figure}

The weather data source is NOAA's Rapid Refresh model RAPv4 \cite{benjamin2016north} with a $13km \times 13km$ high-resolution grid. The \texttt{MetPy} package \cite{may2019metpy}, in the \texttt{Unidata} framework \cite{fulker1997unidata}, is used to capture \texttt{grib2} format data from the NOAA's NCEI THREDDS server. An example is given in \Cref{fig:weather}. We make use of the historical reanalysis archive which had better coverage compared to the operational catalog. Despite this, we observe certain missing hours, in which case the model simply defers to the default SLR rating. The gridded RAP weather dataset contains both the temperature as well as ($v_x$,$v_y$) horizontal and vertical components of wind velocity at different altitudes including 80m, which we choose as a close approximation to the height of transmission-system overhead conductors.

\subsection{Network model data} 
This paper utilizes the 2000-bus synthetic network model for ERCOT via the \texttt{PowerSimData} package \cite{xu2020us, Electric42:online}. System capacities and generator cost assumptions are calibrated by \cite{xu2020us} to represent the system as of 2020; the model also provides hourly time series for demand $P_d$ and renewable generation $P_{max}$. Unit $P_{min}$ parameters are set to zero to prevent several low-load hours from otherwise becoming infeasible.

\subsection{Computational platform and workflow}
\begin{figure}[htbp]
\centerline{\includegraphics[width=\columnwidth]{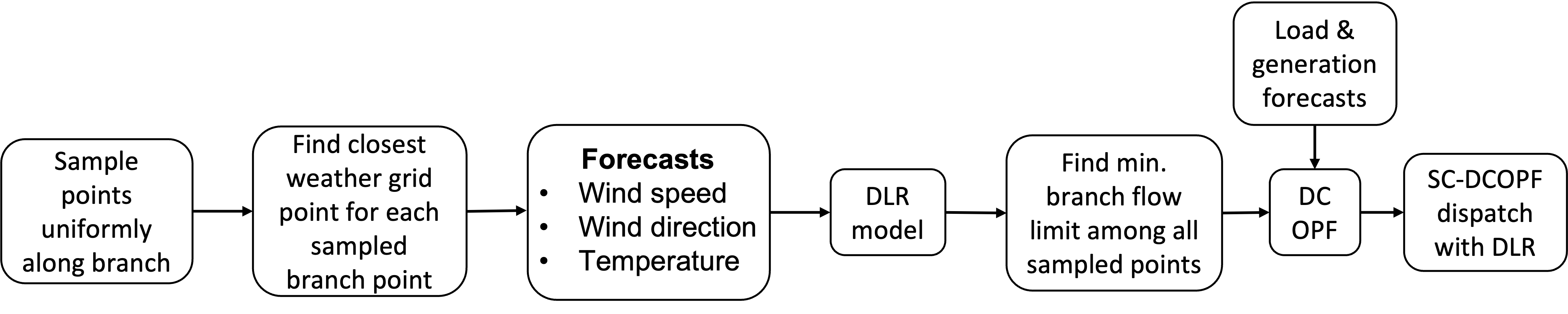}}
\caption{Overall work flow for DLR estimation and dispatch.}
\label{fig:workflow}
\end{figure}

The overall workflow used for the DLR estimation and security-constrained dispatch is shown in \Cref{fig:workflow}. 

The Julia package \texttt{PowerModels.jl} is used to compute the PTDF matrix and solve base case DCOPF cases \cite{coffrin2018powermodels}, and \texttt{PowerSystems.jl} is used to calculate the LODF matrix \cite{lara2021powersystems}. The \texttt{JuMP} optimization framework \cite{DunningHuchetteLubin2017} in the \texttt{Julia} language \cite{bezanson2017julia} is used with the \texttt{Gurobi} solver \cite{gurobi}. 
High-performance computing on MIT Supercloud \cite{reuther2018interactive} (using 2 CPU nodes with 48 cores each) enables one full-year scenario to be solved in about 1 hour, i.e. roughly a 48x speedup. 

\subsection{DLR multiplier factors} 
\begin{figure}[htbp]
     \begin{subfigure}[b]{0.49\columnwidth}
         \includegraphics[width=\columnwidth]{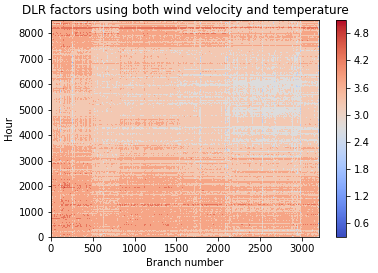}
        \caption{DLR distribution.}
        \label{fig:dlr_dist}    
     \end{subfigure}
     \begin{subfigure}[b]{0.49\columnwidth}
         \includegraphics[width=\columnwidth]{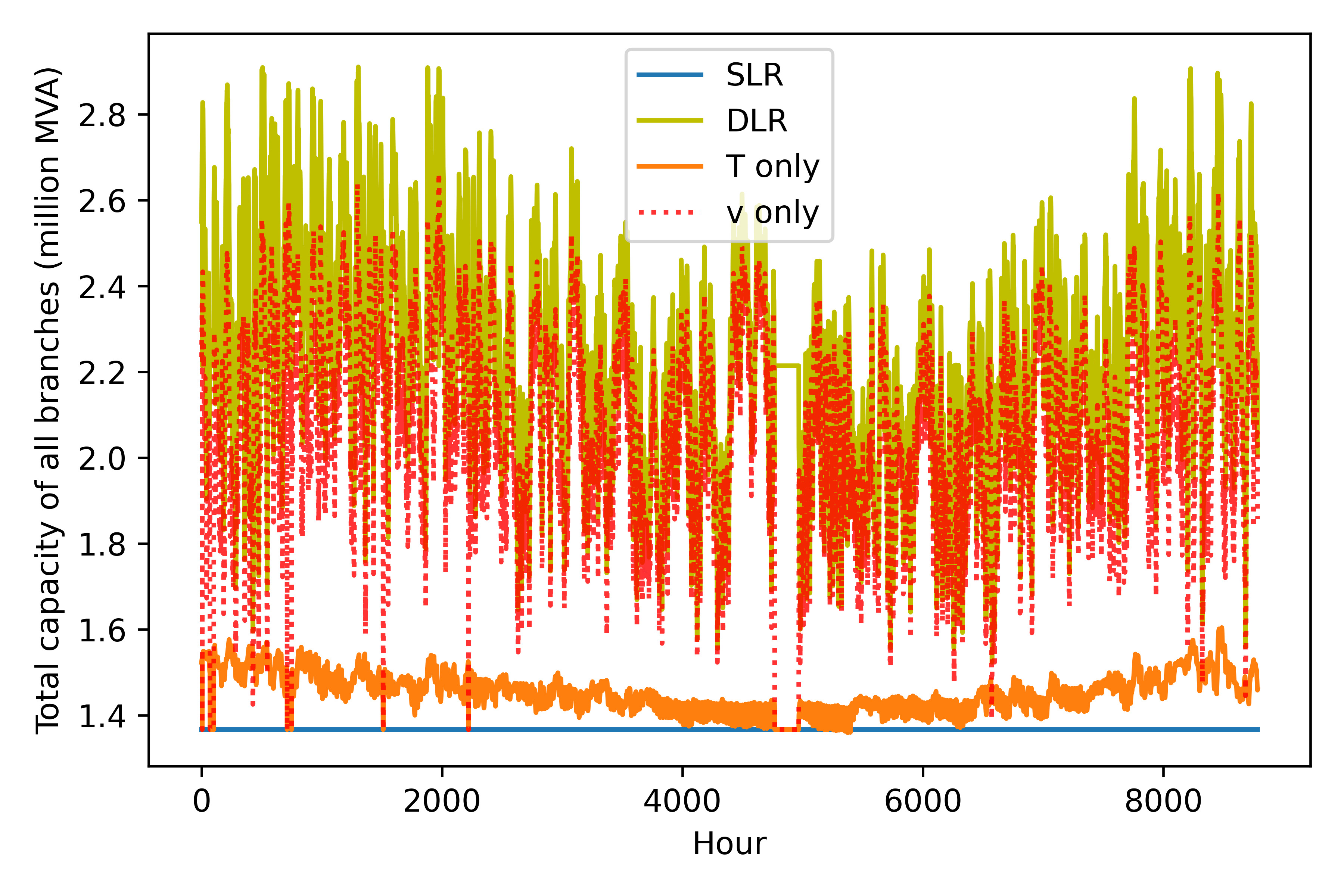}
        \caption{Total system capacity.}
        \label{fig:total_cap}
     \end{subfigure}
     \caption{Summary of DLR results.}
\end{figure}

The DLR multiplier factors obtained from our simulations using a conservative normal conductor temperature of $T_C = 100 \degree C$ and $\phi^{SLR} = 0 \degree$ are shown here. The temporal and spatial distribution of the DLR factors using both wind and temperature inputs (relative to SLR = 1) is shown in \Cref{fig:dlr_dist} across all the sub-100km lines and for each hour in 2016. We see significant variability in the DLR factors and a high correlation among DLR values across different branches and locations for the same hour, due to similar weather conditions. 



The total transmission capacity summed over all branches is shown as a timeseries in \Cref{fig:total_cap}, for different cases with (i) SLR, (ii) AAR using only temperature ($T$ only), (iii) DLR using only wind speed and direction $v$, and (iv) DLR using both $v$ and $T$. These show that including the wind speed and direction in the DLR calculation significantly increases the capacity above the SLR and even the AAR cases.

\subsection{Sensitivity of DLR to assumed parameters}
\begin{figure}[htbp]
    \centering
    \includegraphics[width=0.49\columnwidth]{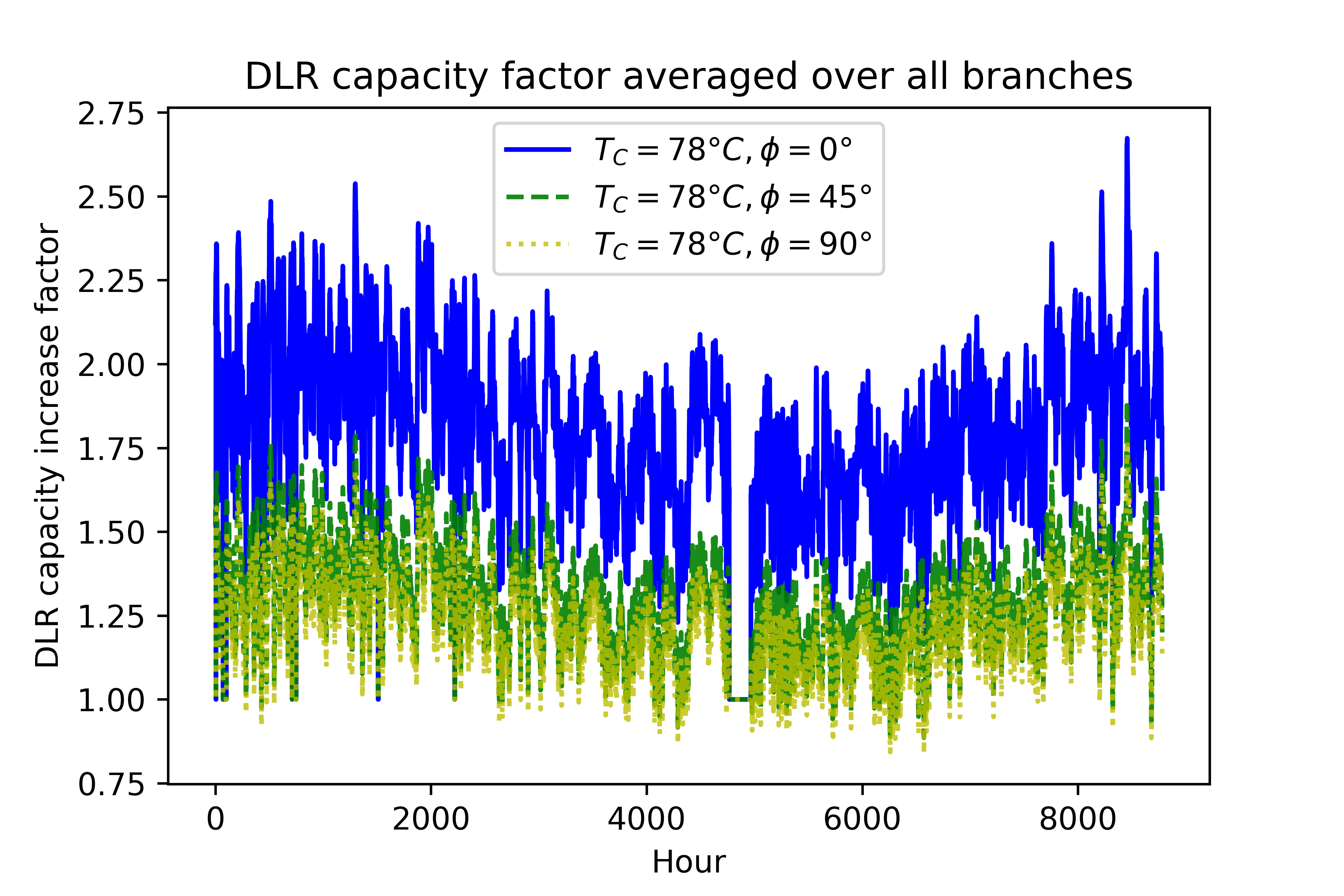}
    \includegraphics[width=0.49\columnwidth]{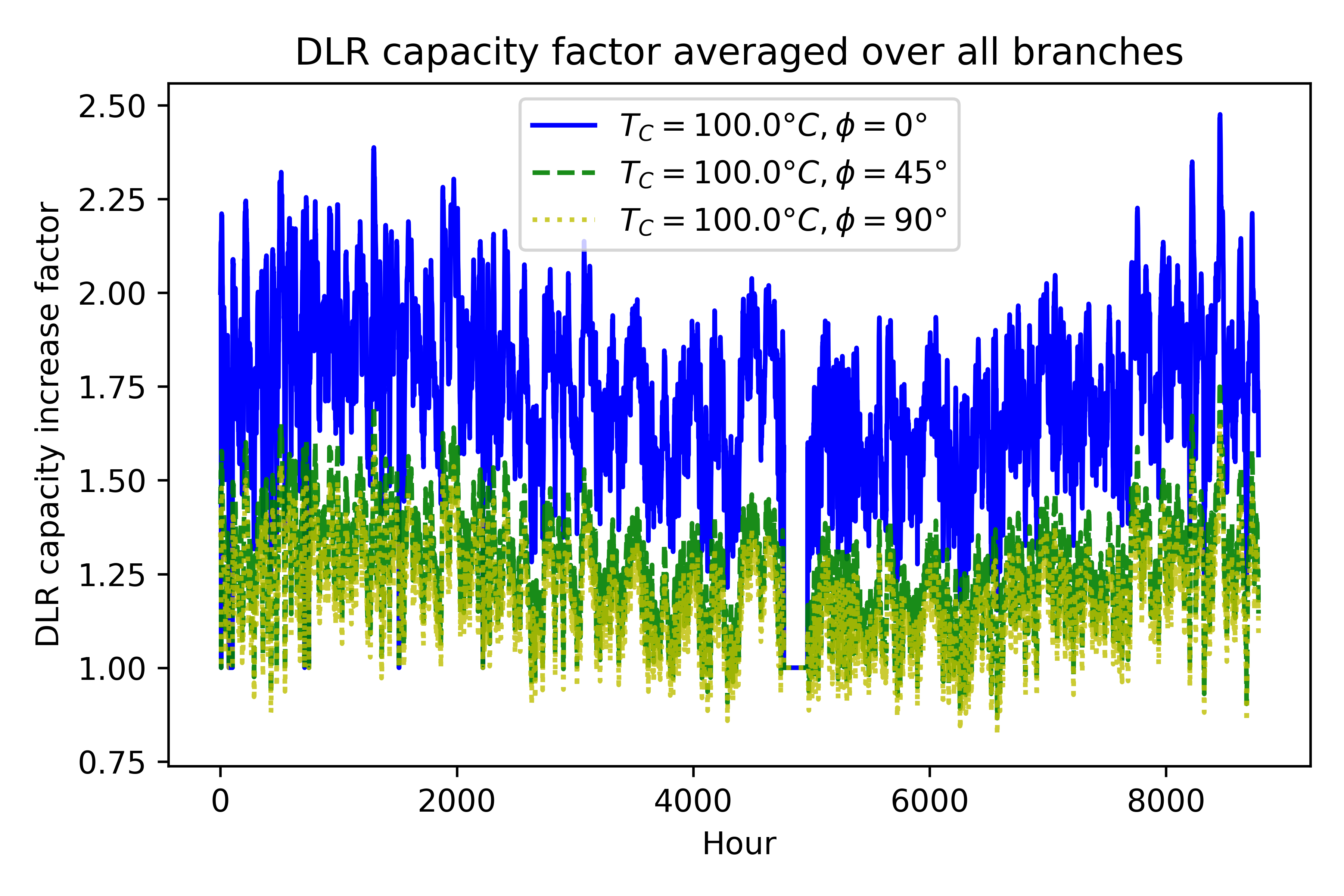}
    \caption{Branch-averaged DLR capacity factor increase values $\eta$ over time, under different parameter values.}
    \label{fig:sensitivity}
\end{figure}

The DLR estimation is also repeated for a few different cases using different values for key parameters including the limiting conductor temperature $T_C$ and the wind angle $\phi^{SLR}$ assumed by transmission system operators for their SLR calculations. A summary of these estimates is shown in \Cref{tab:sensitivity} and \Cref{fig:sensitivity}. These clearly show that when more conservative values are used for the parameters, i.e., higher $T_C$ or lower $\phi^{SLR}$, the higher the resulting DLR capacity increase factors.
\begin{table}[htbp]
\centering
\setlength\tabcolsep{5.5pt}
\begin{tabular}{@{}cccccccccc@{}}
\toprule
$T_C \; [\degree C]$     & 78   & 78   & 78   & 100  & 100  & 100  & 110  & 110  & 110  \\
$\phi^{SLR} \; [\degree]$ & 0    & 45   & 90   & 0    & 45   & 90   & 0    & 45   & 90   \\ \midrule
\textbf{Mean $\eta$}               & 1.79 & 1.30 & 1.23 & 1.73 & 1.26 & 1.19 & 1.71 & 1.25 & 1.18 \\ \bottomrule
\end{tabular}
\caption{DLR capacity factors averaged over branches and all hours in 2016, under different assumed parameter values.}
\label{tab:sensitivity}
\end{table}

\subsection{Effects of DLR on security-constrained economic dispatch}

\subsubsection{Impacts on renewables curtailment and emissions} 
\begin{figure}[htbp]
\centerline{\includegraphics[width=\columnwidth]{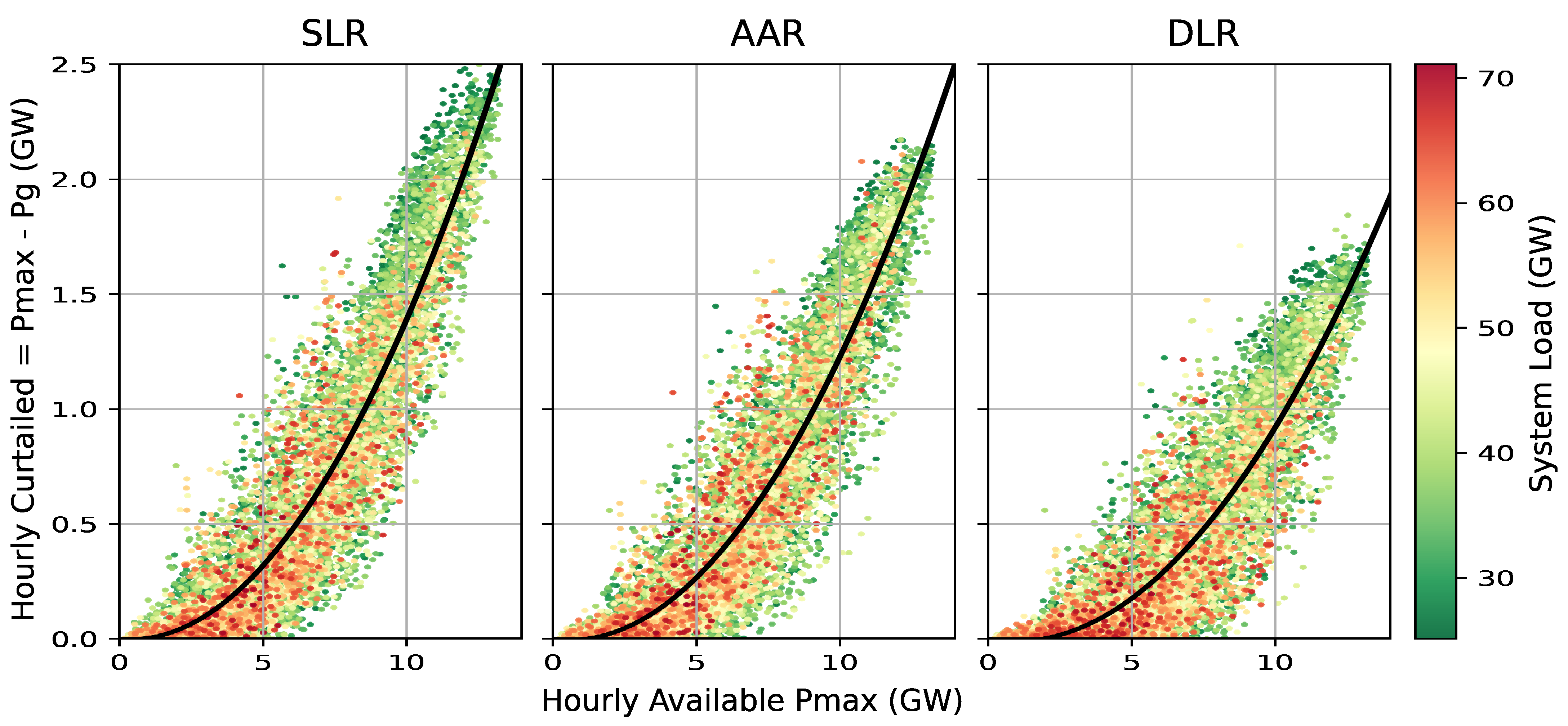}}
\caption{Hourly wind curtailment versus available wind resources: The slope of the positive association diminishes when switching from SLR to AAR to DLR. The 2nd-order polynomial best-fit lines are shown for clarity. Color of scatter plot points correspond to system load levels. The darker green points towards the top right illustrate the contribution of low-load, high-wind hours toward wind curtailment.}
\label{fig:curtail_scatter}
\end{figure}

\begin{figure}[htbp]
\centerline{\includegraphics[width=\columnwidth]{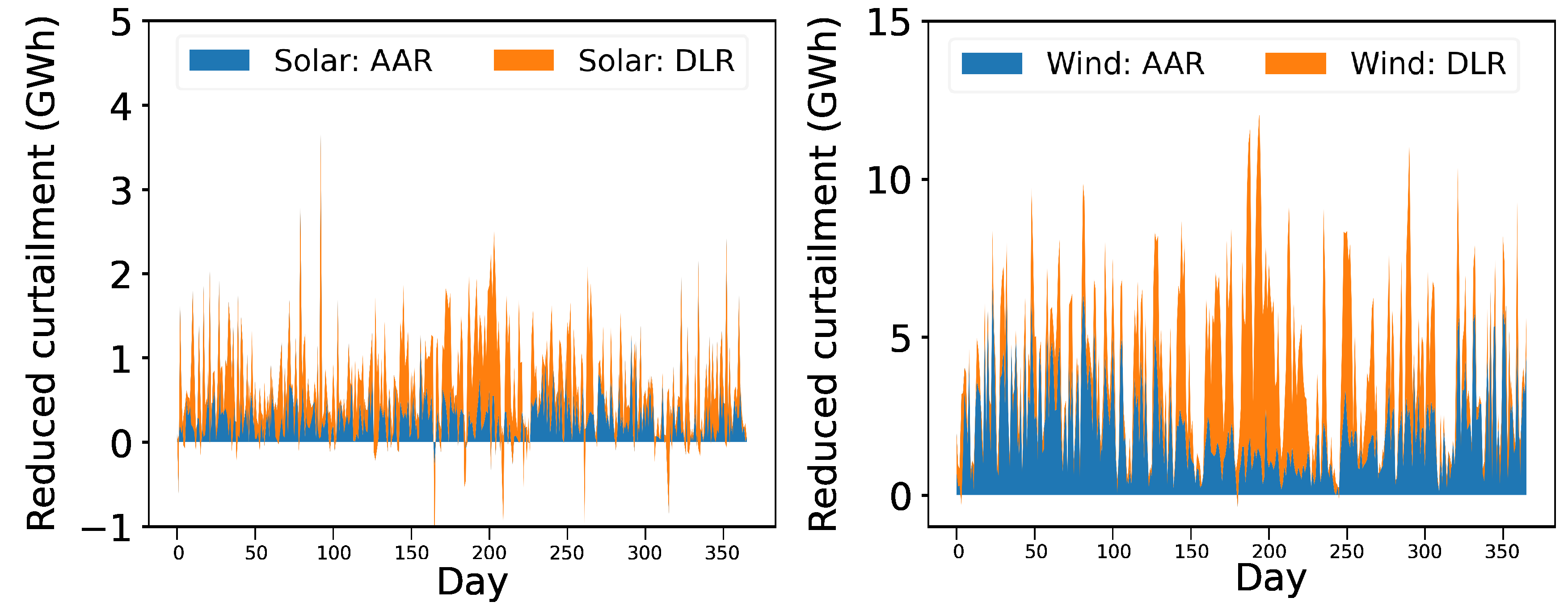}}
\caption{Alleviation of renewable curtailment across time: daily total reduction in curtailment for solar (top) and wind (bottom) generation versus SLR.}
\label{fig:curtail_plot}
\end{figure}

Overall, adopting flexible transmission capacity increases renewable utilization and decreases fossil fuel usage, shown in \Cref{tab1}. Using AAR partially alleviates curtailment of solar and wind production, while introducing DLR further alleviates curtailment by an additional 4.1\% and 2.8\% for solar and wind respectively. In \Cref{fig:curtail_scatter}, the vertical decrease of wind curtailment during DLR is the largest for the highest wind $P_{max}$ hours, illustrating how the coincidence of wind speed with both wind availability and thermal cooling helps to reduce curtailment especially in those hours with high wind resources (and thus high initial SLR curtailments). \Cref{fig:curtail_plot} shows the additional benefits towards curtailment mitigation when using DLR beyond what is achievable under AAR, and demonstrates that both solar and wind improvements occur throughout all seasons albeit with high daily variability.

Whereas using AAR reduces coal power plants usage by about 1\% and total emissions by about 0.6\%, using DLR further reduces coal power plants dispatch by another 3.3\% accompanied by a 1.4\% decrease in emissions. Natural gas generation's slight emission increase under DLR is more than offset by coal power plants' larger magnitude decrease and emissions factor. 

\begin{table}[htbp]
\caption{Generation dispatch results (annual totals).}
\begin{center}
\begin{tabular}{|c|c|c|c|c|c|}
\multicolumn{1}{c}{} & \multicolumn{4}{c}{\textbf{Generation} (TWh)} & \multicolumn{1}{c}{\textbf{Emissions}}\\ 
\hline
Rating type & {Solar} & {Wind}& {NG}& {Coal} & {CO$_2$ (MMT)} \\
\hline
SLR & 4.68 & 50.61 & 182.37 & 67.59 & 143.64\\
AAR & 4.84 & 51.41 & 182.13 & 66.86 & 142.81\\
DLR & 5.04 & 52.86 & 182.69 & 64.65 & 140.81\\
\hline
\multicolumn{1}{c}{} & \multicolumn{4}{c}{\textbf{\% Differences vs. SLR}} \\ 
\hline
Rating type & {Solar} & {Wind}& {NG}& {Coal} & {CO$_2$} \\
\hline
AAR & 3.5\% & 1.6\% & (0.1\%) & (1.1\%) & (0.6\%)\\
DLR & 7.8\% & 4.4\% & 0.2\%  & (4.3\%) & (2.0\%)\\
\hline
\multicolumn{6}{l}{Using coal and natural gas (NG) emission factors from \cite{USEIA:online}.}\\
\multicolumn{6}{l}{MMT = million metric tons. Numbers in parenthesis are negative.}
\end{tabular}
\label{tab1}
\end{center}
\end{table}

\begin{table}[htbp]
\caption{System cost results.}
\begin{center}
\begin{tabular}{|c|r|r|}
\multicolumn{1}{c}{} & \multicolumn{2}{c}{\textbf{Annual Total Costs} (Million \$)}  \\
\hline
Rating Type & System total cost & Congestion cost               \\
\hline
SLR & \$ 10,278 & \$ 1,011 \\
AAR & \$ 9,921 & \$ 654 \\
DLR & \$ 9,502 & \$ 235 \\
Uncongested & \$ 9,267 & \$ 0\\
\hline
\multicolumn{1}{c}{} & \multicolumn{2}{c}{\textbf{Savings vs. SLR}}    
\\ \hline
Rating Type & (Million \$) & (\% of potential) \\
\hline
AAR         & \$                   (356) &                 (35.2\%) \\
DLR         & \$                   (776) &                 (76.8\%) \\
\hline
\multicolumn{3}{l}{Uncongested = no transmission constraints.}\\
\multicolumn{3}{l}{System total cost = sum of objective values.}\\
\multicolumn{3}{l}{Congestion = system total cost $-$ uncongested total cost.}
\end{tabular}
\label{tab2}
\end{center}
\end{table}

\subsubsection{Impacts on system economics}

Table \ref{tab2} summarizes system costs when using different line rating types. Compared to an annual cost of \$10.3 billion when using SLR, using AAR and DLR reduces system costs by \$356 million (3.5\%) and \$776 million (7.5\%) respectively. A copperplate economic dispatch simulation is run for all hours by removing transmission constraints, in order to estimate the uncongested total costs and decompose the congestion component in \Cref{tab2}. Compared to this lower-bound benchmark with infinite transmission capacity, AAR and DLR achieve 35\% and 77\% respectively of the total potential congestion savings.

\begin{figure}[htbp]
\centering
\includegraphics[width=\columnwidth]{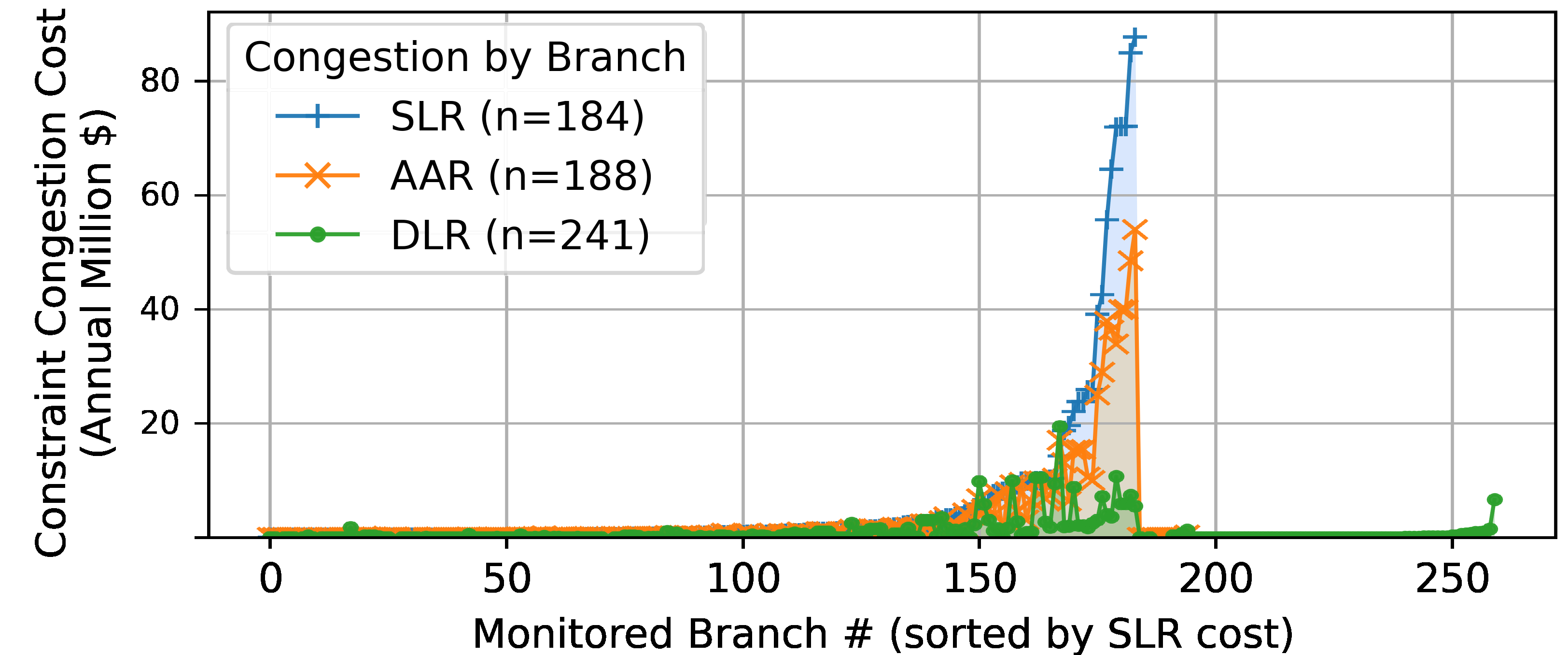}
\caption{Distributions of binding constraints' costs, grouped by monitored branch, sorted by SLR then DLR costs.}%
\label{fig:cong_line}%
\end{figure}

\begin{figure}[htbp]
\centering
\includegraphics[width=2.8cm]{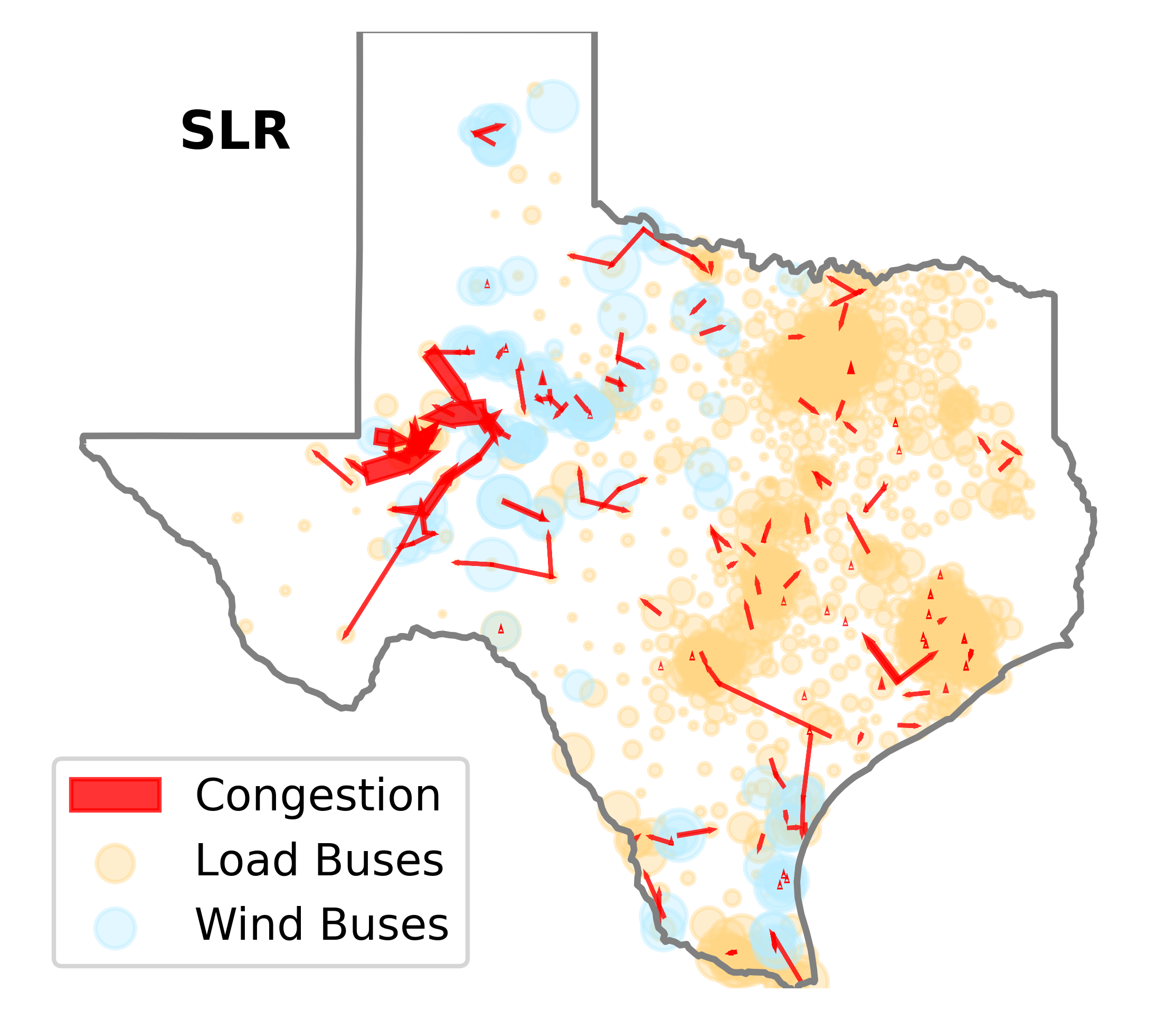}
\includegraphics[width=2.8cm]{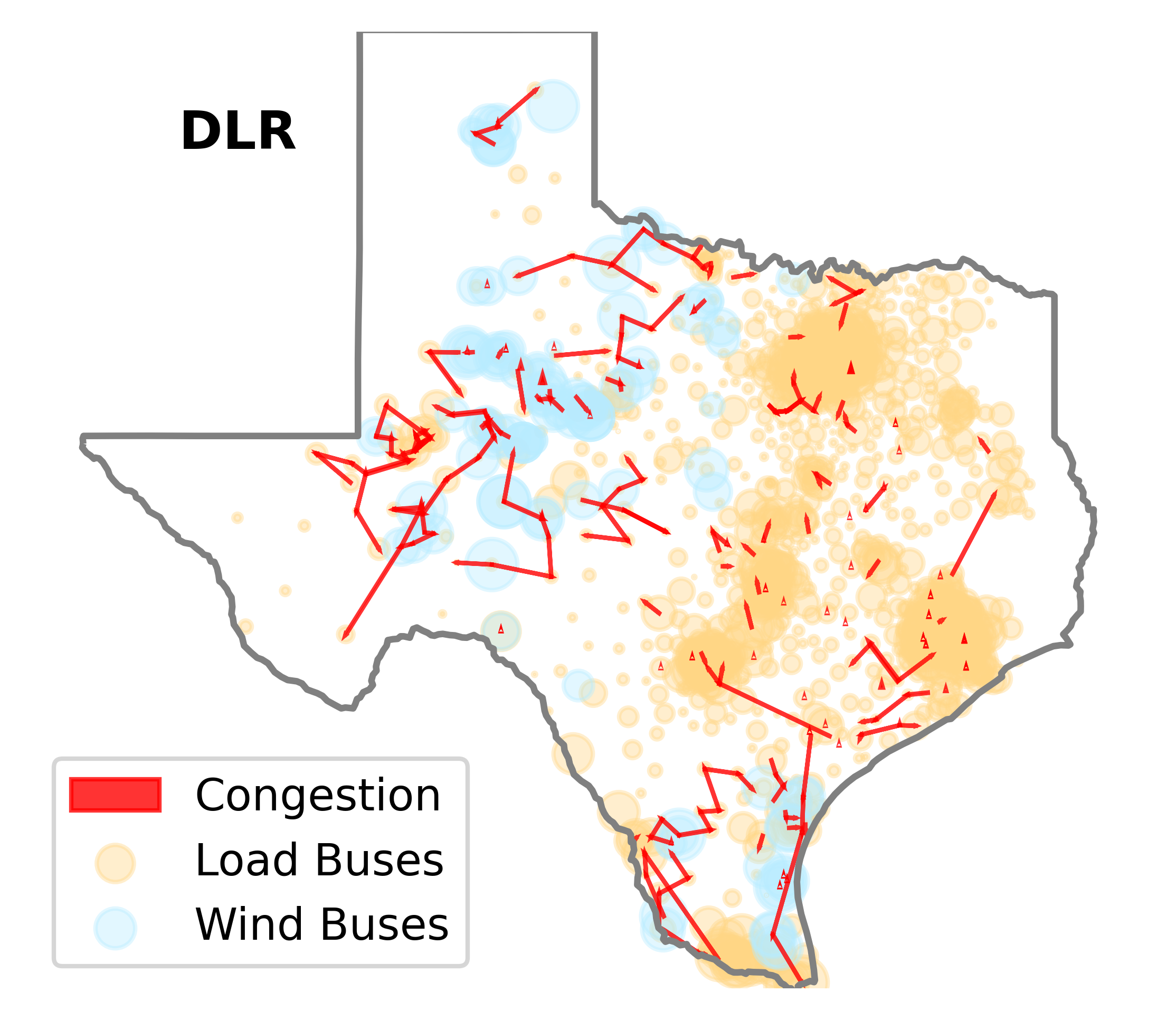}
\includegraphics[width=2.8cm]{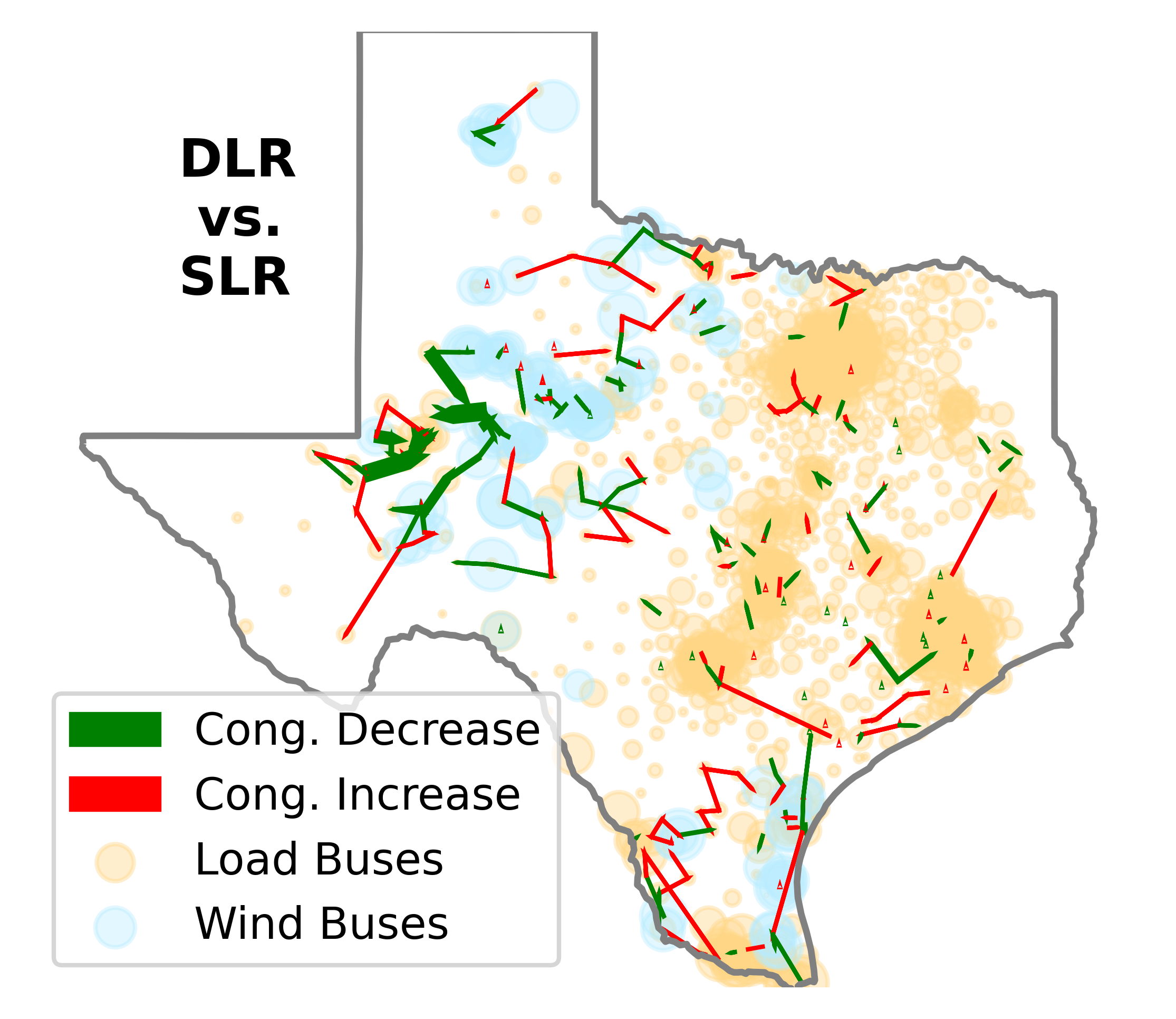}
\caption{Geospatial distribution of network congestion costs by monitored branches, when using SLR and DLR, as well as the delta due to switching from SLR to DLR. Binding branches are shown as lines with widths proportional to each branch's total congestion cost across the study year. Load bus (orange) sizes correspond to average nodal load $P_d$, while wind generation bus (blue) sizes correspond to installed capacity.}%
\label{fig:maps}%
\end{figure}

\Cref{fig:cong_line} shows how the costliest congested lines under SLR are partially relieved by AAR, and further relieved by implementing DLR. The ordering is sorted based on the costs under SLR, so the shapes of the AAR and DLR distributions show how the congestion change is not monotonic. Specifically, while DLR consistently resolves the most severe $\sim$20\% of the SLR and AAR constraints, some of the constraints which appeared moderate under SLR actually worsens when using DLR (as evidenced by the green line rising near the middle of the plot). In addition, adopting AAR and DLR has actually increased the number of binding constraints from 184 originally, to 188 and 241 respectively. However, these new ``next in line" constraints result in relatively low shadow prices, which do not offset the overall congestion relief effects. \Cref{fig:maps} shows the spatial distribution of the top binding constraints during the year from the numerical results. Many of the original constraints in far western Texas near adjacent wind resources are significantly resolved by DLR (green lines in the bottom ``DLR vs. SLR" map). Meanwhile, the additional red lines illustrate the locations of the aforementioned, less severe ``next in line" constraints.

\subsection{Caveats and limitations}
Despite its potential transformative benefits, there may be several practical challenges to widespread adoption of DLR by utilities and system operators. Firstly, the highly weather-dependent DLR values are quite volatile which may cause operational difficulties for dispatch and market clearing due to forecasting uncertainty. Further, more validation of the ampacities predicted by DLR may be needed in order to meet stringent reliability requirements. Deploying DLR would also require software implementation updates to incorporate multi-dimensional weather inputs, compared to the simpler temperature lookup tables for AAR.

\section{Conclusion}
In this paper, security-constrained DCOPF was solved at hourly resolution on a synthetic network model of ERCOT, when using three different sets of line ratings: SLR, AAR, and DLR. The AAR estimates roughly approximate ERCOT's current operational practice, while the DLR levels represent a potential next-step improvement that also incorporates atmospheric wind velocity and direction data. The numerical results suggest that adopting such a DLR approach would yield benefits that are twice or more compared to the level achieved by AAR, when compared against a baseline of SLR. The approximate doubling was seen when comparing metrics including renewable curtailment mitigation (solar 2.2x, wind 2.75x), carbon emission mitigation (3.3x), as well as economic cost savings (2.2x) driven by the alleviation of key network constraints. While further analysis is needed, DLR may offer significant transmission system benefits beyond AAR.


\bibliographystyle{IEEEtran}
\bibliography{references,refs}

\end{document}